\documentclass[aps,prl,reprint,superscriptaddress]
{revtex4-2}

\usepackage{amsmath}
\usepackage{amsfonts}
\usepackage{hyperref}
\usepackage{graphicx}
\usepackage{nicefrac}

\begin{document}

\title{Electron spin secluded inside a bottom-up assembled standing metal-molecule nanostructure}

\author{Taner Esat}
\email[Corresponding author: ]{t.esat@fz-juelich.de}
\affiliation{Peter Gr\"unberg Institute (PGI-3), Forschungszentrum J\"ulich, 52425 J\"ulich, Germany}
\affiliation{J\"ulich Aachen Research Alliance (JARA), Fundamentals of Future Information Technology, 52425 J\"ulich, Germany}

\author{Markus Ternes}
\affiliation{Peter Gr\"unberg Institute (PGI-3), Forschungszentrum J\"ulich, 52425 J\"ulich, Germany}
\affiliation{J\"ulich Aachen Research Alliance (JARA), Fundamentals of Future Information Technology, 52425 J\"ulich, Germany}
\affiliation{Institute of Physics II B, RWTH Aachen University, 52074 Aachen, Germany}

\author{Ruslan Temirov}
\affiliation{Peter Gr\"unberg Institute (PGI-3), Forschungszentrum J\"ulich, 52425 J\"ulich, Germany}
\affiliation{J\"ulich Aachen Research Alliance (JARA), Fundamentals of Future Information Technology, 52425 J\"ulich, Germany}
\affiliation{Institute of Physics II, University of Cologne, 50937 Cologne, Germany}

\author{F.~Stefan Tautz}
\affiliation{Peter Gr\"unberg Institute (PGI-3), Forschungszentrum J\"ulich, 52425 J\"ulich, Germany}
\affiliation{J\"ulich Aachen Research Alliance (JARA), Fundamentals of Future Information Technology, 52425 J\"ulich, Germany}
\affiliation{Experimental Physics IV A, RWTH Aachen University, 52074 Aachen, Germany}

\date{\today}

\begin{abstract}
Artificial nanostructures, fabricated by placing building blocks such as atoms or molecules in well-defined positions, are a powerful platform in which quantum effects can be studied and exploited on the atomic scale. Here, we report a strategy to significantly reduce the electron-electron coupling between a large planar aromatic molecule and the underlying metallic substrate. To this end, we use the manipulation capabilities of a scanning tunneling microscope (STM) and lift the molecule into a metastable upright geometry on a pedestal of two metal atoms. Measurements at millikelvin temperatures and magnetic fields reveal that the bottom-up assembled standing metal-molecule nanostructure has an $S = \nicefrac{1}{2}$ spin which is screened by the substrate electrons, resulting in a Kondo temperature of only $291 \pm 13$\,mK. We extract the Land\'e $g$-factor of the molecule and the exchange coupling $J\rho$ to the substrate by modeling the differential conductance spectra using a third-order perturbation theory in the weak coupling and high-field regimes. Furthermore, we show that the interaction between the STM tip and the molecule can tune the exchange coupling to the substrate, which suggests that the bond between the standing metal-molecule nanostructure and the surface is mechanically soft.
\end{abstract}

\maketitle
On the way to spin qubits based on single atoms or molecules, it is essential to minimize the interaction with the environment, since the latter leads to decoherence \cite{Willke2022}. The scanning tunneling microscope (STM) is an ideal tool to study quantum properties of nanoscale structures, because it not only allows the magnetic states of individual atoms and molecules to be read out \cite{Heinrich2004} and coherently controlled \cite{Yang2019, Willke2021, Veldman2021}, but also enables the environment to be changed directly. The ability to arrange atoms and molecules on surfaces with atomic precision allows for the fabrication and study of unprecedented artificial nanostructures \cite{Khajetoorians2019, Yang2021}. Moreover, the STM can be used to fabricate multiple absolutely identical qubits \cite{He2019} from individual atoms and molecules, which can also be arranged and coupled with each other as desired. Compared to mesoscopic qubits, the structural control down to the atomic level may offer advantages.

Magnetic atoms and molecules with degenerate ground states on metallic surfaces typically show the Kondo effect: the spin degree of freedom is quenched at temperatures below a characteristic Kondo temperature $T_K$ by the formation of a many-electron singlet state with the  electrons of the bath \cite{Kondo1964, Wilson1975}. Because $T_K$ depends directly on the coupling with the metal, the Kondo effect itself can be used as a gauge of the interaction with the environment. The strong hybridization of the d-orbitals of  magnetic atoms with states of the metal substrate leads to $T_K$ of typically 40 - 300\,K \cite{Ternes2008}. For magnetic molecules, on the other hand, Kondo temperatures of only a few Kelvin have been observed on metal surfaces \cite{Zhang2013, Garnier2020, Frauhammer2021}, which can be explained by the weaker hybridization of the molecular orbitals with the substrate, or the shielding of the magnetic atoms by the surrounding ligands of the molecule. However, long relaxation times $T_1$ of several hundred nanoseconds up to days \cite{Paul2017, Natterer2017} and dephasing times $T_2$ in the nanosecond range \cite{Yang2019, Willke2021, Veldman2021} were so far only achieved for atoms and molecules that were decoupled from the metallic surface by an atomically thin insulating layer. The presence of the decoupling layer has also resulted in a significant reduction of $T_K$ to a few Kelvin for magnetic atoms \cite{Otte2008, Yang2018}.

\begin{figure}
\includegraphics{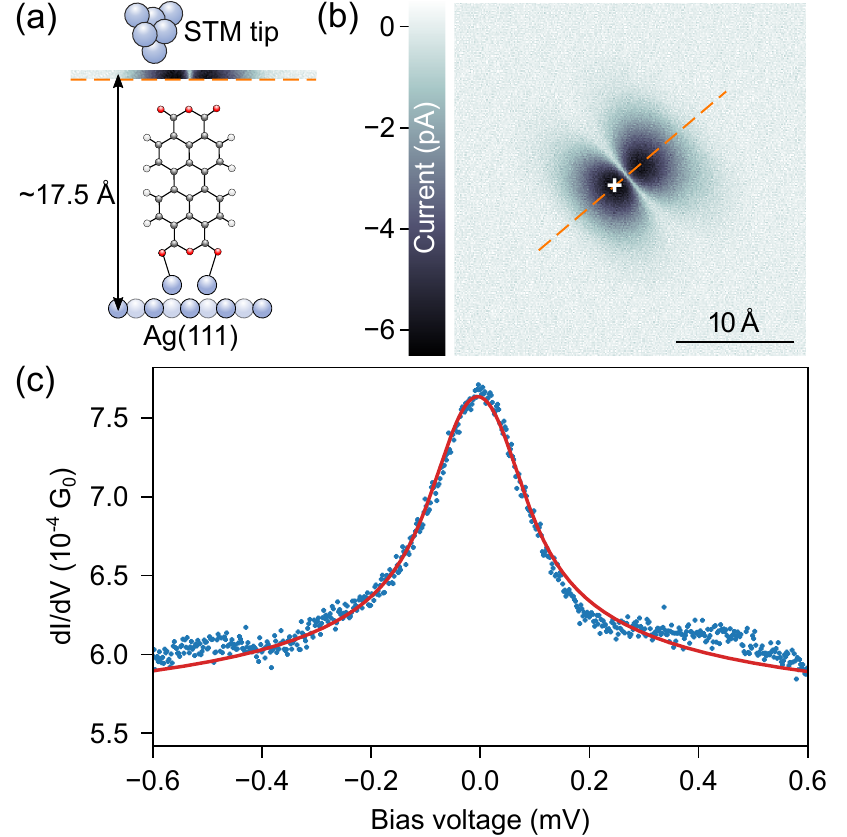}
\caption{(a) Schematic view of a standing PTCDA + 2Ag nanostructure on the Ag(111) surface, including the STM tip above the molecule. The bar shows the tunneling current $I_T$ measured above the standing nanostructure at constant height. The white, grey and red spheres indicate hydrogen, carbon and oxygen atoms, respectively, of PTCDA. (b) Constant-height STM image above a standing nanostructure recorded at a tip height of $z \simeq 17.5$\,\AA\:above the surface. The bias voltage was $V = -50$\,mV. The white cross marks the location where the $dI/dV$ conductance spectra were measured. The molecular plane is indicated by the dashed orange line. (c) $dI/dV$ conductance spectrum (blue) on a standing metal-molecule nanostructure measured at $T = 30$\, mK and $B=0$\,T ($V_{\mathrm{mod}} = 50$\,µV). The tip was stabilized at $I_T = 45$\,pA and $V = -1$\,mV. The red curve shows the fit based on the Frota function (see text for details). The spectrum is shown in units of the conductance quantum $G_0 = \frac{2e}{h}$.}
\label{Fig:fig1}
\end{figure}

In this work, we show that exploiting the third dimension for the bottom-up assembly of standing metal-molecule nanostructures offers an alternative approach to tune the coupling with the metallic substrate. Specifically, we show that for a single 3,4,9,10-perylenetetracarboxylic dianhydride (PTCDA) in the standing configuration on a pedestal of two Ag adatoms (Fig.~\ref{Fig:fig1}a), both the interaction with the Ag(111) substrate is drastically reduced compared to the flat-lying PTCDA and the coupling with the metal substrate can be tuned by stretching the bond between the molecule and surface, utilizing the attraction between the STM tip and the molecule. We report the fabrication of an $S=\nicefrac{1}{2}$ spin nanostructure based on this strategy, with a very weak coupling to the underlying substrate, resulting in a $T_K$ of only $291 \pm 13$\,mK --- to our knowledge, the smallest $T_K$ ever measured on a metallic substrate using STM. Comparably low $T_K$ have so far only been found in mesoscopic quantum dots \cite{Cronenwett1998, WielS2000}.

The Ag(111) surface was prepared in ultra-high vacuum (UHV) by repeated Ar$^{+}$ sputtering and heating at 800\,K. A small coverage of PTCDA molecules was evaporated onto the clean Ag(111) surface at room temperature from a custom-built Knudsen cell. After evaporation, the sample was flashed at 480\,K for 2\,min and then cooled down to 100 K and transferred to the STM. All experiments were performed in the J\"ulich Quantum Microscope \cite{Esat2021}, a millikelvin scanning tunneling microscope which uses the adiabatic demagnetization of electronic magnetic moments in a magnetocaloric material to reach temperatures in the range between 30\,mK and 1\,K. In this instrument, $B$ fields of up to 8\,T perpendicular to the sample surface can also be applied. Differential conductance ($dI/dV$) spectra were measured using conventional lock-in techniques with the STM feedback loop switched off and an AC modulation amplitude $V_{\mathrm{mod}} = 20 - 100$\,µV and frequency $f_{\mathrm{mod}} = 187$\,Hz. The PtIr tip was treated in-situ by applying controlled voltage pulses and indentations into the clean silver surface until the spectroscopic signature of the Ag(111) surface state appeared.

The standing PTCDA + 2Ag nanostructure was fabricated on the Ag(111) surface in three steps by controlled manipulation with the tip of the STM as described in Ref.~\cite{Esat2018}. First, two single Ag atoms were attached to the two carboxylic oxygens on one side of the flat-lying molecule by lateral manipulation with the tip. Then, one of the carboxylic oxygens on the opposite side was contacted and the PTCDA molecule was pulled up on a curved trajectory until it stood upright. The tip was then moved straight up until the bond between the molecule and the tip broke, leaving the molecule in the standing position on the two Ag adatoms \cite{Esat2018}. The stability of the standing metal-molecule nanostructure arises from the balance between local covalent interactions and nonlocal long-range van der Waals forces \cite{Knol2022, Arefi2022}.

In constant-height STM images, the standing metal-molecule nanostructure can be recognized by two features that are distributed symmetrically around the plane of the molecule (dashed orange line in Fig.~\ref{Fig:fig1}b) and separated by a nodal plane perpendicular to the latter (Fig.~\ref{Fig:fig1}b). It is interesting to note that the node of the $\pi$-orbital in the plane of the molecule could not be resolved. The two features coincide with the positions where the interaction with the tip is most pronounced \cite{Esat2018}. In the standing configuration, there is only a weak overlap between the wave functions of the metallic surface and the lowest unoccupied molecular orbital (LUMO) of PTCDA, because the lobes of the molecular $\pi$-orbital are oriented perpendicular to the plane of the molecule. This allows the standing nanostructure to function as a quantum dot and coherent field emitter \cite{Esat2018}. 

\begin{figure}
\includegraphics{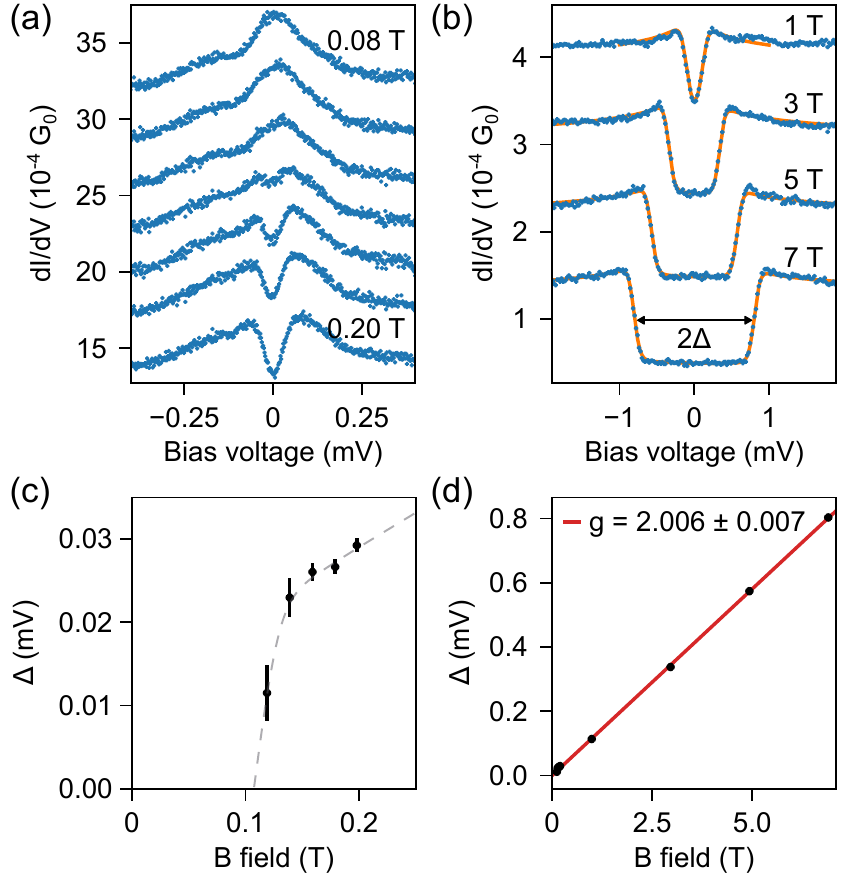}
\caption{(a)-(b) $dI/dV$ spectra (blue) on a standing metal-molecule nanostructure, measured at different $B$ fields at $T \simeq 50$\,mK (setpoints in panel (a) were $I_T = 100$\,pA, $V = -10$\,mV, $V_{\mathrm{mod}} = 100$\,µV and $I_T = 100$\,pA, $V = -1$\,mV, $V_{\mathrm{mod}} = 20$\,µV in panel(b)). In panel (a), the B field was changed in steps of $20$\,mT. The orange curves in panel (b) show the fits based on perturbation theory (see text for details). The spectra are vertically displaced for clarity. (c)-(d) The Zeeman splitting $\Delta$ extracted from the $dI/dV$ spectra as a function of $B$. The gray dashed line in panel (c) serves as a guide for the eye. The red curve in panel (d) shows the linear fit for the Zeeman splitting.}
\label{Fig:fig2}
\end{figure}

At mK temperatures, a peak at zero bias is evident in the $dI/dV$ spectrum measured on a standing metal-molecule nanostructure (Fig.~\ref{Fig:fig1}c). In fact, at these low temperatures we additionally observe a dip at zero bias due to the dynamical Coulomb blockade (DCB) \cite{Ast2016}. We have thus corrected all $dI/dV$ spectra on the standing nanostructure for the DCB dip (see Supplementary Material). Previous studies have hinted that the LUMO of the standing metal-molecule nanostructure must contain a single unpaired electron \cite{Esat2018, Temirov2018}. Therefore, it is plausible to assume that the zero-bias peak originates from the Kondo effect, in which the spin of this localized electron is screened by itinerant substrate electrons. To verify this, we measured $dI/dV$ spectra at different $B$ fields. Already at $B \approx 100 - 120$\,mT a Zeeman splitting of the zero-bias peak is discernable (Fig.~\ref{Fig:fig2}a). At higher $B$ fields, the Kondo effect is completely quenched and the spectrum is dominated by the symmetric steps arising from inelastic spin-flip excitations (Fig.~\ref{Fig:fig2}b). 

To extract the precise energy of the Zeeman splitting $\Delta$, we calculated the numerical derivative of the $dI/dV$ spectra and fitted the peak positions with a Gaussian (see Supplementary Material). As shown in Figs.~\ref{Fig:fig2}c and d, the energies of the spin-flip excitations scale linearly with the external $B$ field. Only close to the critical field $B_C$, which is required to initially split the Kondo resonance, the Zeeman energy rises noticeably faster with increasing $B$ field. To extract the Land\'e factor $g$, we consider only the data points at $B$ fields $\geq 1$\,T (Fig.~\ref{Fig:fig2}b) since the Kondo effect in the strong coupling regime leads to renormalization of the $g$-factor. A linear fit of the form $\Delta = g \mu_B B$ for the Zeeman effect, where $\mu_B$ is the Bohr magneton, yields a Land\'e factor $g = 2.006 \pm 0.007$. By interpolating the data points at low B fields, we obtain $B_C= 108 \pm 5$\,mT for the critical field. Using the relation \cite{Costi2000} 
\begin{equation}
B_C = \frac{1}{2} \frac{k_B T_K}{g \mu_B},
\end{equation} 
valid for temperatures $T < 0.25\,T_K$, this gives an estimate of $291 \pm 13$\,mK for the Kondo temperature $T_K$. 

An independent estimate of the Kondo temperature $T_K$ can be obtained from the width of the Kondo resonance (Fig.~\ref{Fig:fig1}c). However, it should be noted that this is only a rough estimate, since the width of the Kondo resonance is related to $T_K$ by a non-universal scaling constant \cite{Esat2015}. To extract its width, we fitted the Kondo resonance with a Frota line shape \cite{Frota1992}
\begin{equation}
\rho(E)_{\text{Frota}} = \Re\sqrt{\frac{i \Gamma_K}{E-E_0 + i \Gamma_K}} \text{.}
\end{equation}
Additional broadening effects due to the Fermi distribution and the modulation amplitude were taken into account. The best fit then yields a width of $\Gamma_K \simeq 43\,\mu$V and thus a $T_K = \Gamma_K (2\pi \times 0.103)/k_B \simeq 320$\,mK \cite{Frota1992}, in good agreement with the above estimate from the $B$ field dependence. We attribute the features in the $dI/dV$ spectrum at approximately $\pm 0.25$\,mV and $\pm 0.55$\,mV (Fig.~\ref{Fig:fig1}c and Supplementary Material) to either molecular vibrations or frustrated translations of the standing metal-molecule nanostructure \cite{Knol2022}. Note that a strong electron-vibrational coupling can also lead to a further decrease of $T_K$ \cite{Eickhoff2020}.

The low Kondo temperature of the standing nanostructure in conjunction with the low base temperature and high energy resolution of our mK STM enable us to quantitatively describe the interaction of the localized spin with its environment, also as a function of temperature. For this purpose, we employ the Anderson-Appelbaum model \cite{Appelbaum1966, Anderson1966, Appelbaum1967} and calculate the tunneling conductance from the Kondo Hamiltonian in a perturbative approach that includes processes up to third order in the exchange interaction $J$ \cite{Ternes2015}. The model allows tunneling electrons to interact with the localized spin via spin-spin ($t_{TS} \,\hat{\mathbf{\sigma}}_t\cdot \hat{\mathbf{S}}$) or potential scattering ($t_{TS} \,U$). Here $t_{TS}$ is the matrix element for a transition from the tip to the molecule or vice versa, and $\hat{\mathbf{\sigma}}_t$ and $\hat{\mathbf{S}}$ are the spin operators of the tunneling and localized electrons, respectively. In addition, the model takes into account the spin-spin exchange scattering between the electrons of the substrate and the localized spin ($J\rho \,\hat{\mathbf{\sigma}}_s\cdot \hat{\mathbf{S}}$), where $\rho$ denotes the substrate's electron density at the Fermi energy and $\hat{\mathbf{\sigma}}_s$ the spin operator of itinerant electrons in the substrate. This approach provides the correct description under the following conditions: the magnetic impurity is predominantly coupled to one of the electrodes (here the substrate), the system is in equilibrium (limit of small bias voltages), and the system is in the weak coupling limit ($T \gtrsim T_K$) or high-field regime ($B \gtrsim k_B T_K$). We performed least-square fits and extracted the dimensionless coupling strength $J\rho$ between the substrate and the localized electron and its Land\'e factor $g$.

\begin{figure}
\includegraphics{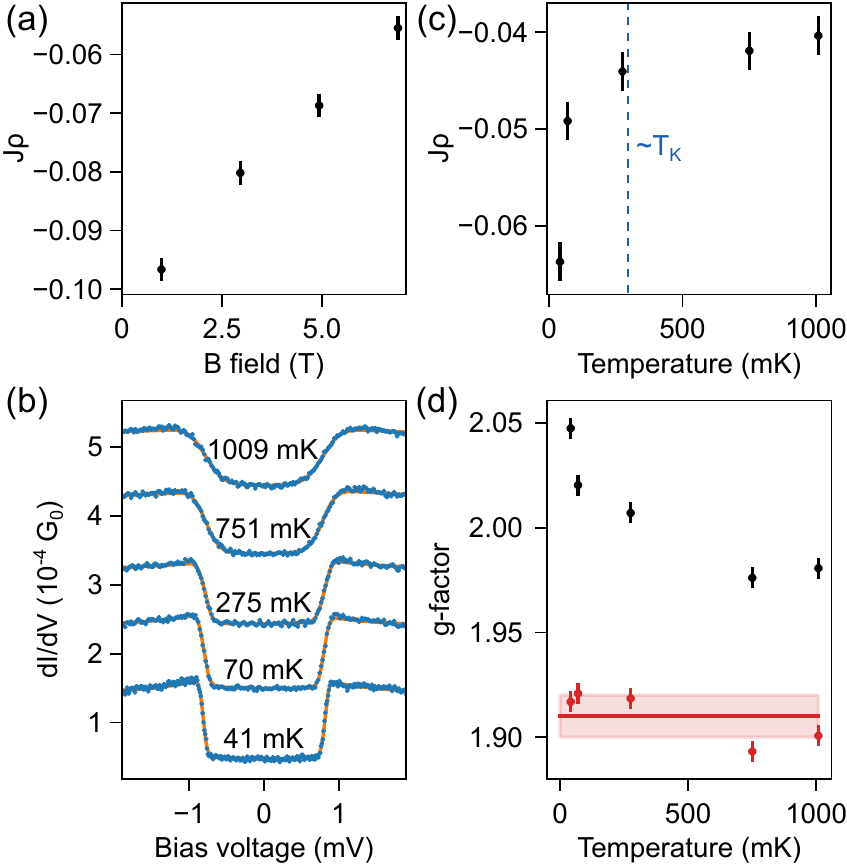}
\caption{(a) Coupling strength $J\rho$ as extracted from the fits in Fig.~\ref{Fig:fig2}b as a function of $B$ field. (b) $dI/dV$ spectra (blue) on a standing metal-molecule nanostructure, measured at different temperatures in an external field $B = 7$\,T ($I_T = 100$\,pA, $V = -10$\,mV, $V_{\mathrm{mod}} = 100$\,µV). The orange curves show the fits based on perturbation theory (see text for details). The spectra are vertically displaced for clarity. (c) Coupling strength $J\rho$ as extracted from the fits as a function of temperature. The dashed blue line indicates the Kondo energy scale $T_K$ as determined from the $B$-field data of Fig.~\ref{Fig:fig2}. (d) Land\'e $g$-factor estimated from the fits in panel (b) (black) and the effective $g$-factor $g_\mathrm{eff}$ after taking into account renormalization effects due to the exchange interaction (red). The red line illustrates a linear fit and the red shaded area the corresponding confidence interval.}
\label{Fig:fig3}
\end{figure}

Before we focus on the temperature dependence, we first examine the influence of the B field on $J\rho$ at $50$\,mK (Fig.~\ref{Fig:fig2}b). As the B field increases, we see a decrease in the height of the peak structure on top of the steps originating from spin-flip excitations. Since those peak heights are proportional to $J\rho$ \cite{Ternes2015}, this indicates that $|J\rho|$ decreases with increasing B field. This is clearly seen in Fig.~\ref{Fig:fig3}a, where the fitted $J\rho$ is plotted versus B field. It may at first sight seem surprising that we still observe a coupling between the localized and itinerant spins at high B fields. This behaviour is, however, in good agreement with numerical renormalization group (NRG) calculations for an $S = \nicefrac{1}{2}$ Kondo impurity at finite temperatures and $B$ fields \cite{Costi2000}, in which it was shown that the intensity of the split Kondo resonance varies even if $\mu_B B / k_B T_K \gg 1$, which corresponds to the present situation. In other words, we have even at high $B$ fields access to bias driven Kondo correlations whose gradual emergence at decreasing temperatures drives $|J\rho|$ up. This behavior can be readily observed by looking at the temperature-dependent data for constant $B = 7$\,T (Fig.~\ref{Fig:fig3}b). The fits reveal that $|J\rho|$ increases with decreasing temperature (Fig.~\ref{Fig:fig3}c). For $B=0$, such an increase of $|J\rho|$ would signal the progressive breakdown of the perturbation approach, yielding a divergence of $J\rho$ and the crossover into the Kondo singlet as a new ground state \cite{Kondo1964, Wilson1975}. However, here we are in the high-field regime and therefore will not reach the Kondo ground state even in the limit $T \rightarrow 0$. We note that the temperature range in which the perturbation theory starts to collapse (Fig.~\ref{Fig:fig3}b) agrees very well with the Kondo energy scale of $291 \pm 13$\,mK which was derived from the $B$-field behaviour at low temperatures $T<T_K$.

For the fitted Land\'e factor $g$ in Fig.~\ref{Fig:fig3}d we also see a strong decrease with increasing temperature. This can be attributed to the energy renormalization \cite{Ternes2015}. Taking this into account, we obtain an effective gyromagnetic factor of $g_{\mathrm{eff}} = g(T)\times(1 + J\rho(T)) \approx 1.91 \pm 0.01$, which has no temperature dependence. Note that the deviation of the obtained g-factors from the $B$-field-dependent measurements and from the perturbative approach is $\lesssim 6$\%.

\begin{figure}
\includegraphics{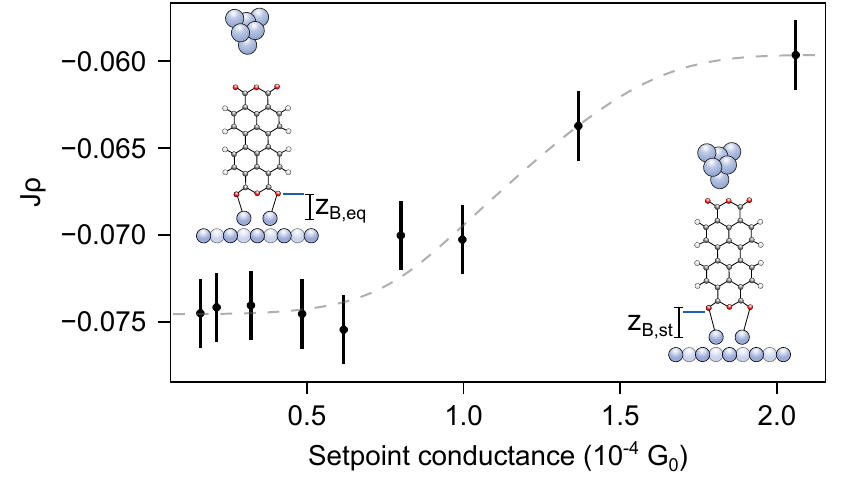}
\caption{Coupling strength $J\rho$ as extracted from the fits of $dI/dV$ spectra on a standing metal-molecule nanostructure that were measured for different setpoint conductances at $T \simeq 45$\,mK and an external field of $B = 7$\,T. The tip was initially stabilized at $I_T = 100$\,pA and $V = -6$\,mV and then moved up by $1$\,\AA\:in steps of $0.1$\,\AA. The gray dashed line serves as a guide for the eye. Insets show schematically how the PTCDA molecule is pulled up by the tip.}
\label{Fig:fig4}
\end{figure}

Having demonstrated the sensitivity to changes of $J\rho$, we now explore the possibility to tune the exchange coupling mechanically. It was shown before that the standing metal-molecule nanostructure is susceptible to attractive forces from the tip, enabling controlled tilts, translations and rotations \cite{Esat2018}. If it was possible to tune the vertical distance of the standing nanostructure from the substrate with attractive forces from the STM tip, it might also be possible to tune the exchange coupling $J\rho$. To explore this possibility, we measured $dI/dV$ spectra on the standing metal-molecule nanostructure at $B = 7$\,T and $T \simeq 45$\,mK for different setpoint conductances, corresponding to different distances between the tip and the molecule. In Fig.~\ref{Fig:fig4} the fitted $J\rho$ are plotted as a function of setpoint conductances $G$. For $G \leq 0.6 \cdot 10^{-4} G_0$, the coupling is constant at $(J\rho)_\text{eq} \simeq -0.075$. With increasing $G$, corresponding to decreasing tip-molecule distances, $|J\rho|$ decreases and reaches $(J\rho)_\text{st} \simeq -0.060$ for the highest $G$. Measurements at even smaller distances (higher setpoints) are not feasible, because the resulting larger tunnel currents frequently induce sudden $30^{\circ}$ rotations of the standing nanostructure around its vertical axis. Since we know that there are attractive forces acting between the molecule and the tip \cite{Esat2018, Knol2022}, we interpret the decreasing $|J\rho|$ as the result of an increased distance of the standing metal-molecule nanostructure from the surface. Under the assumption that the exchange interaction scales exponentially with the bond distance $z_B$ \cite{Yang2019a}, $J(z_B) \propto \exp(-z_B / d_\mathrm{ex})$, the vertical relaxation $\Delta z_B$ of the bond between the standing molecule and substrate surface can be estimated. For a typical decay length of the exchange interaction,  $d_\mathrm{ex} \simeq 0.4$\,\AA\, \cite{Yang2019a}, we obtain $\Delta z_B = d_\mathrm{ex} \ln(J_\text{eq} / J_\text{st}) \simeq 0.09$\,\AA \, between the smallest and the largest $G$ in Fig.~\ref{Fig:fig4}.

In conclusion, we have shown that in the standing configuration the exchange coupling between PTCDA within the assembled nanostructure and the Ag(111) substrate is strongly reduced, if compared to the flat-lying molecule. At $B=0$ we observed a Kondo resonance with a width of only $\Gamma_K \simeq 43$\,µV at an experimental temperature of $T = 30$\,mK. $B$-field-dependent measurements showed that the standing metal-molecule nanostructure is an $S=\nicefrac{1}{2}$ system with a critical field of $B_C= 108 \pm 5$\,mT. This corresponds to a Kondo temperature of only $T_K = 291 \pm 13$\,mK. Furthermore, we demonstrated that, using attractive forces exerted by the STM tip, it is possible to tune the exchange coupling between the localized spin in the nanostructure and the substrate. The combination of the small exchange coupling and the softness of the surface bond against vertical distortions makes the standing metal-molecule nanostructure an interesting candidate for STM-based electron spin resonance (STM-ESR) experiments. For STM-ESR experiments on individual atoms and molecules \cite{Yang2019, Willke2021, Yang2019a, Baumann2015, Seifert2020, Yang2018, Veldman2021, Yang2021, Natterer2017, Willke2019, Weerdenburg2021, Seifert2020a, Willke2022}, two important requirements have to be met \cite{Lado2017, Seifert2020}: first, a sufficiently small coupling between the object to be investigated and the substrate, in order to reach long relaxation and dephasing times, and second, the possibility to drive with a high-frequency electric field applied to the STM tip mechanical oscillations of the object in the inhomogeneous $B$ field of the tip, the latter being produced by a magnetic atom at the tip apex. Although a significant reduction of the interaction between the atomic or molecular object of interest with the metal substrate has been achieved on different atomically thin insulating layers \cite{Loth2010, Paul2017}, ESR signals in the STM have been observed, somewhat surprisingly, mainly on a bilayer of magnesium oxide (MgO) film on Ag(001) surfaces \cite{Yang2019, Willke2021, Yang2019a, Baumann2015, Seifert2020, Yang2018, Veldman2021, Yang2021, Natterer2017, Willke2019, Weerdenburg2021, Seifert2020a, Willke2022} and recently for the first time on two-monolayer NaCl films on Cu(100) \cite{Kawaguchi2022}. In this situation, the standing metal-molecule nanostructure may be a promising specimen: its spin is more weakly coupled to the substrate than that of Cu atoms on MgO/Ag(001), which are ESR active \cite{Yang2018}, and the tip-induced displacement is about an order of magnitude larger than the displacements required for ESR-STM \cite{Lado2017, Seifert2020}. It should be noted, however, that for ESR-STM a dynamic displacement driven by the high-frequency electric field is required, whereas in the present experiment we so far only tested the response to static forces between the molecule and the tip. But due to the strong molecular polarizability of PTCDA \cite{Temirov2018}, we anticipate that the high-frequency electric field may have a similar effect. In upcoming experiments we will therefore determine whether the dynamic displacement is sufficiently large, and whether relaxation times are sufficiently long for STM-ESR experiments. Finally, we note that standing metal-molecule nanostructure can also be prepared on the tip \cite{Wagner2015, Temirov2018, Wagner2019}. If it was indeed STM-ESR capable, the standing nanostructure could therefore be employed as a magnetic field sensor on the atomic scale \cite{Verlhac2019, Garnier2020}, in addition to being a sensor of electric surface potentials, as which it has already been used \cite{Wagner2015, Wagner2019}. 

\begin{acknowledgments}
We thank Frithjof B. Anders (TU Dortmund) for fruitful discussions. The authors acknowledge financial support from the German Federal Ministry of Education and Research through the funding program 'quantum technologies - from basic research to market', under Q-NL (project number 13N16032). M.T. acknowledges funding by the Heisenberg Program (TE 833/2-1) of the Deutsche Forschungsgemeinschaft (DFG). F.S.T. acknowledges funding by the DFG through SFB 1083 "Structure and Dynamics of Internal Interfaces" (223848855-SFB 1083).
\end{acknowledgments}

\end{document}